\documentclass[pre,twocolumn,a4paper,superscriptaddress,floatfix]{revtex4}
\usepackage{graphicx}
\usepackage{bbm}
\usepackage[all]{xy}
\usepackage{amsmath}
\usepackage{amssymb}
\usepackage{epstopdf}

\def\be{\begin{equation}}
\def\ee{\end{equation}}
\newcommand{\bq}{\begin{eqnarray}}
\newcommand{\eq}{\end{eqnarray}}
\newcommand{\bes}{\begin{subequations}}
\newcommand{\ees}{\end{subequations}}

\def\ben{\begin{eqnarray}}
\def\een{\end{eqnarray}}
\def\ba{\begin{array}}
\def\ea{\end{array}}

\begin{document}
\newcommand{\half}{{\textstyle\frac{1}{2}}}
\allowdisplaybreaks[3]
\def\a{\alpha}
\def\b{\beta}
\def\g{\gamma}\def\G{\Gamma}
\def\d{\delta}\def\D{\Delta}
\def\ep{\epsilon}
\def\et{\eta}
\def\z{\zeta}
\def\t{\theta}\def\T{\Theta}
\def\l{\lambda}\def\L{\Lambda}
\def\m{\mu}
\def\f{\phi}\def\F{\Phi}
\def\n{\nu}
\def\p{\psi}\def\P{\Psi}
\def\r{\rho}
\def\s{\sigma}\def\S{\Sigma}
\def\ta{\tau}
\def\x{\chi}
\def\o{\omega}\def\O{\Omega}
\def\k{\kappa}
\def\pa {\partial}
\def\ov{\over}
\def\br{\\}
\def\ud{\underline}

\newcommand\lsim{\mathrel{\rlap{\lower4pt\hbox{\hskip1pt$\sim$}}
    \raise1pt\hbox{$<$}}}
\newcommand\gsim{\mathrel{\rlap{\lower4pt\hbox{\hskip1pt$\sim$}}
    \raise1pt\hbox{$>$}}}
\newcommand\esim{\mathrel{\rlap{\raise2pt\hbox{\hskip0pt$\sim$}}
    \lower1pt\hbox{$-$}}}

\title{Unified paradigm for interface dynamics}

\author{P.P. Avelino}
\email[Electronic address: ]{ppavelin@fc.up.pt}
\affiliation{Centro de F\'{\i}sica do Porto, PT-4169-007 Porto, Portugal}
\affiliation{Departamento de F\'{\i}sica da Faculdade de Ci\^encias
da Universidade do Porto, PT-4169-007 Porto, Portugal}
\affiliation{Departamento de F\'{\i}sica, Universidade Federal da Para\'{\i}ba
58051-970 Jo\~ao Pessoa, Para\'{\i}ba, Brazil}
\author{R. Menezes}
\email[Electronic address: ]{rmenezes@dce.ufpb.br}
\affiliation{Centro de F\'{\i}sica do Porto, PT-4169-007 Porto, Portugal}
\affiliation{Departamento de F\'{\i}sica da Faculdade de Ci\^encias
da Universidade do Porto, PT-4169-007 Porto, Portugal}
\affiliation{Departamento de Ci\^encias Exatas, Universidade Federal da Para\'{\i}ba, 58297-000 Rio Tinto PB, Brazil}
\affiliation{Departamento de F\'{\i}sica, Universidade Federal de Campina Grande, 58109-970, Campina Grande, Para\'{\i}ba, Brazil}
\author{J.C.R.E. Oliveira}
\email[Electronic address: ]{jespain@fe.up.pt}
\affiliation{Centro de F\'{\i}sica do Porto, PT-4169-007 Porto, Portugal}
\affiliation{Departamento de Engenharia F\'{\i}sica da Faculdade de Engenharia
da Universidade do Porto, PT-4200-465 Porto, Portugal}

\begin{abstract}

In this paper we develop a common theoretical framework for the dynamics of thin featureless interfaces. We explicitly demonstrate that the same phase field and velocity dependent one-scale models characterizing the dynamics of relativistic domain walls, in a cosmological context, can also successfully describe, in a friction dominated regime, the dynamics of 
nonrelativistic interfaces in a wide variety of material systems. We further show that a statistical version of von Neumann's law applies in the case of scaling relativistic interface networks, implying that, although relativistic and nonrelativistic interfaces have very different dynamics, a single simulation snapshot is not able to clearly distinguish the two regimes. We highlight that crucial information is contained in the probability distribution function for the number of edges of domains bounded by the interface network and explain why laboratory tests with nonrelativistic interfaces can be used to rule out cosmological domain walls as a 
significant dark energy source.

\end{abstract} 
\maketitle

\section{Introduction}

Phase field models (PFM) have been applied with success to describe the interface dynamics of various nonrelativistic systems,  including foams  \cite{1992JPCM....4.1867G,foams} and grain growth \cite{2006PhRvE..74f1605K} (see also \cite{2002AdPhy..51..481B, 2008AdPhy..57....1E} for recent reviews of PFM). In PFM sharp interfaces are modeled as diffuse boundaries, with a thickness much smaller than the curvature radius, separating regions characterized by different values of the order parameters. In the simplest cases with nonconserved order parameters, such as ideal grain growth, cellular patterns bounded by an interface network tend to evolve toward configurations with less interfacial energy (or equivalently, less surface area). This leads to the gradual elimination of entire domains, resulting in the increase of the characteristic scale $L$ with physical time $t$ as $L \propto t^{1/2}$ both for two- and three-dimensional networks \cite{1994PhRvB..5015752C,1998PhRvL..80.5228M,2006PhRvE..74f1605K}.

In these systems the velocity $v$ is proportional to the mean curvature of the interface at each point. This is the crucial ingredient 
behind von Neumann's law \cite{vonNeumann:1952:DRC} for the evolution of the area of individual domains in two-dimensional interface networks with (free) Y-type junctions: $d a_n/dt \propto (n-6)$, where $a_n$ is the area of a domain with $n$ edges. Since the average number of edges of an interface network with Y-type junctions is $6$, the only possible static configuration consistent with von Neumann's law is a hexagonal lattice. Note that if the interfaces have a finite width, von Neumann's law may not apply to individual domains but is reasonably obeyed on average \cite{0295-5075-15-5-002}. Extended versions of von Neumann's law exist in both two \cite{1992PhRvL..69..208A} and three dimensions \cite{PhysRevLett.72.420,2001PhRvL..86.2685H,nature1} (see also Ref. \cite{2009ConPh..50..199W}). Although other laws, such as Lewis's law and Aboav's law, have also been used in the characterization of classical patterns \cite{2009ConPh..50..199W,0305-4470-39-23-004,PhysRevE.47.4037}, in this paper we shall focus on von Neumann's law, which is more directly relevant for the evolution of an interface network.

On the other hand, the dynamics of relativistic interface networks, considering various possible types of defect junctions, has been studied in detail in a cosmological context, where domain walls have been proposed in the past as an interesting dark energy candidate \cite{Bucher:1998mh}. In Refs. \cite{PinaAvelino:2006ia,Avelino:2008ve}, the authors have used large and accurate sets of field theory numerical simulations in combination with semi-analytical analysis to provide very strong evidence for a no-frustration conjecture, which invalidates domain walls as a viable dark energy candidate. 

The main aim of this paper is the development of a unified theoretical paradigm for interface dynamics which includes both relativistic and nonrelativistic systems in a unified framework. We set up a velocity-dependent one-scale (VOS) model and use it to describe the evolution of the characteristic length $L$ and velocity ${\bar v}$ of thin, featureless interface networks both in the relativistic and nonrelativistic limits. We show that the same models which describe the dynamics of cosmic relativistic interfaces also account, in a friction dominated regime, for the dynamics of cellular patterns in many material systems, such as soap froths, lipid monolayers, and metals. Then we compare single simulation snapshots of high resolution relativistic and nonrelativistic interface network simulations, discussing the cosmological implications of our findings, in particular for dark energy. We also demonstrate that von Neumann's law still applies in the relativistic regime, albeit only on average. In this paper we use natural units with $c=\hbar=1$ and the Einstein summation convention.

\section{Unified framework}

We shall consider a multicomponent field theory Lagrangian with a discrete symmetry. For simplicity let us choose a model with a scalar field multiplet $\phi_e$ described by the Lagrangian 
\be
\mathcal{L}=X-V(\phi_e)\,,
\ee 
where $X=-\partial_\mu \phi_e \, \partial^\mu \phi_e / 2$ and $V(\phi_e)$ is the potential (with the exception of  Eq. (\ref{ideal}), in the following we shall omit the index $e$). This model admits interface solutions if the potential $V(\phi)$ has at least two discrete degenerate minima. Although models described by more general Lagrangians can be considered, that does not introduce additional dynamical degrees of freedom, as long as the 
interfaces are thin and featureless  \cite{Avelino:2008ve,Sousa:2010zz}. Still, one should bear in mind that there are very rich  spectra of geometrical, energetic, and topological network properties associated with different choices of the 
potential $V(\phi)$.

Varying the action,
\be
S=\int d^4x\sqrt{-g}\mathcal{L}\,,
\ee
with respect to the scalar field $\phi$ one obtains the following equation of motion
\be
\frac{1}{\sqrt{-g}}\partial_\mu\left(\sqrt{-g}\partial^\mu\phi \right)=\frac{\partial V}{\partial \phi}\,.
\label{eom1}
\ee
Here $g=\det(g_{\alpha\beta})$ and $g_{\alpha\beta}$ is the metric tensor. In a flat homogeneous and isotropic three-dimensional background the line element is $ds^2=-dt^2+R^2(t){\bf dx} \cdot {\bf dx}$, where $R(t)$ is the scale factor, $t$ is the physical time, and ${\bf x}=(x,y,z)$ are comoving coordinates. In this case, the equation of motion for the scalar field $\phi$, given by  Eq.  (\ref{eom1}), can be written as
\be
\ddot \phi+\frac{\dot \phi}{\ell_{\rm d}}-\nabla^2_{\bf r}  \phi=-\frac{\partial V}{\partial \phi}\,,
\label{eom2}
\ee
where a dot represents a derivative with respect to physical time, $\ell_{\rm d}^{-1}=3H$, $H={\dot R}/R$, ${\bf r}=R(t){\bf x}$, and $\nabla^2_{\bf r}$ is the physical Laplacian. In the presence of frictional forces the characteristic damping length becomes 
$\ell_{\rm d}^{-1}=3H+\ell_{\rm f}^{-1}$, where $\ell_{\rm f}$ is the friction length scale.

It has been shown in Ref. \cite{Sousa:2010zz} that if the thickness of the interface is very small compared to its curvature radius then Eq.  (\ref{eom2}) implies that
\be
{\dot v}+\left(1-v^2\right)\left[\frac{v}{\ell_{\rm d}}-\kappa\right]=0\,,
\label{dyn1}
\ee
where $v$ is the microscopic interface velocity and $\kappa$ is proportional to the mean curvature of the interface at each point.

Assuming that the damping is homogeneous so that $\ell_{\rm d}=\ell_{\rm d}(t)$, the macroscopic equation of motion may 
be written as 
\be
{\dot {\bar v}}+\left(1-{\bar v}^2\right)\left[\frac{{\bar v}}{\ell_{\rm d}}-{\bar \kappa}\right]=0\,,
\label{vos1}
\ee
where
\be
{\bar v}^2= \frac{\int v^2 \rho dV}{\int \rho dV}\,,\qquad {\bar \kappa}=\frac{\int v(1-v^2) \kappa \rho dV}{(1-{\bar v}^2){\bar v}\int \rho dV}\,,
\label{defvk}
\ee
$\rho$ is the interface energy density and $V$ is the physical volume. We have assumed that $\langle v^4 \rangle={\bar v}^4$ (the brackets denote a spatial average) in the derivation of Eq. (\ref{vos1}). This assumption is valid in the relativistic regime up to first order in $(1-v)$ and it has a negligible impact in the nonrelativistic limit. Although it is possible to construct network configurations with the same $\bar v$ but quite different ${\bar \kappa}$, in most physically realistic situations it is sufficient to consider that ${\bar \kappa}={\bar \kappa}({\bar v})$ \cite{Martins:2000cs}. 

Energy-momentum conservation in an expanding homogeneous and isotropic background leads to
\be
{\dot {\bar \rho}}+3H\left({\bar \rho}+{\bar p}\right)=0\,,
\label{enc}
\ee
where ${\bar p}=V^{-1}\int p \, dV $ and ${\bar \rho}=V^{-1}\int \rho \, dV$ are the average pressure and density whose ratio is given by 
the equation of state parameter, $w={\bar p}/{\bar \rho}=-2/3+{\bar v}^2$. Eq. (\ref{enc}) can be generalized in order to account for additional energy loss mechanisms, due to interface collapse and friction:
\be
{\dot {\bar \rho}}+3H\left({\bar \rho}+{\bar p}\right)=-\left(\frac {{\tilde c}{\bar v}}{L}+\frac{{\bar v}^2}{\ell_{\rm f}}\right){\bar \rho}\,.
\label{enc1}
\ee
Here, ${\tilde c}$ is a phenomenological parameter which depends on the specific properties of the network and can be calibrated 
using numerical simulations  \cite{Avelino:2008ve}, $\ell_{\rm f}$ is the friction length scale, and $L$ is the characteristic length of the interface network defined by ${\bar \rho}=\sigma/L$, where $\sigma$ is the interface energy per unit area. Using Eq.  (\ref{enc1}) one obtains 
\be
{\dot L}=HL+\frac{L}{l_d} {\bar v}^2+{\tilde c}{\bar v}\,.
\label{vos2}
\ee
Eqs. (\ref{vos1}) and (\ref{vos2}) constitute the VOS model for domain walls which was proposed in \cite{Avelino:2005kn} and is here derived in detail. In particular the relation between the macroscopic parameter ${\bar \kappa}$ and its microscopic counterpart $\kappa$ has been determined only in the present paper.

Consider the case with $\ell_{\rm f}=\infty$ and $R \propto t^\alpha$ (with $\alpha > 0$). The attractor solution to the evolution Eqs. (\ref{vos1}) and (\ref{vos2}) corresponds to a linear scaling solution with $H^{-1}=t/\alpha$ and
\be
\frac{L}{t}= {\sqrt {\frac{k(k+{\tilde c})}{3 \alpha (1-\alpha)}}}\,, \qquad {\bar v}={\sqrt {\frac{1-\alpha}{3\alpha}\frac{k}{k+{\tilde c}}}}\,, \label{scaling}
\ee
where $k={\bar \kappa} L$ is a dimensionless curvature parameter. A linear scaling solution can be attained only if $\alpha < 1$ ($\bar v \ge 1$ or $L \le 0$ are unphysical). On the other hand,  if $\alpha > 1/4$ a linear scaling solution is possible even with ${\tilde c}=0$. If $\ell_{\rm f}=\infty$ then the interfaces are expected to be relativistic, as long as the expansion is sufficiently slow, except for very particular and unrealistic network configurations, with $k \sim 0$. These results are supported by numerical simulations of cosmological domain wall networks in a frictionless regime \cite{Avelino:2005kn,Avelino:2008ve}. On the other hand, if domain walls were to provide a significant contribution to the dark energy then including friction would not significantly help frustration due to the limited amount of energy with which domain walls can interact conserving energy and momentum \cite{PinaAvelino:2006ia,Avelino:2008ve}. Hence, it is enough to consider the relativistic case with $\ell_{\rm f}=\infty$ in order to probe domain walls as a relevant dark energy candidate.

In the context of phase field models the background is usually fixed ($\alpha=0$ and $H=0$) and the dynamics of the nonrelativistic interfaces is friction dominated.  In this regime $|\dot v| \ll |v/\ell_{\rm f}|$ and Eq.  (\ref{dyn1}) implies the microscopic relation $v = \kappa \ell_{\rm f}$. If the friction is homogeneous and time independent ($\ell_{\rm f}={\rm constant}$) then the interfaces move with a local velocity proportional to the local curvature (note that in the relativistic regime there is no direct relation between the local curvature and velocity). In the nonrelativistic case it is straightforward to find the following attractor solution to the evolution Eqs. (\ref{vos1}) and (\ref{vos2}):
\be
L={\sqrt {2{\tilde c}k\ell_{\rm f} t}} \,, \qquad {\bar v}=\frac{k \ell_{\rm f}}{L} \ll 1\,,
\label{scalingnr}
\ee
which is the usual $t^{1/2}$ scaling for nonrelativistic interfaces, implied by von Neumann' s law  \cite{1992JPCM....4.1867G,foams}.

\section{Numerical results and discussion}

A specific realization of a particular class of multicomponent models which, in the limit of an infinite number of coupled scalar fields, approaches the so-called ideal model, was developed and studied in detail \cite{PinaAvelino:2006ia,Avelino:2008ve} in a cosmological context. It is characterized by a scalar field multiplet with $N_\phi$ real scalar fields and a scalar field potential with $N_\phi+1$ equivalent minima, equidistant in configuration space,
\begin{equation}
V \propto \sum_{j=1}^{N_\phi+1} r_j^2 \left(r_j^2 - r_0^2\right)^2\ {\rm with}\ 
r_j^2=\sum_{i=1}^{N_\phi} (\phi_i - p_{{i}_{j}})^2\,, \label{ideal}
\end{equation}
where the $p_{{i}_{j}}$ are the $N_\phi+1$ coordinates of the vacua of the potential, chosen to be the vertices of an ($N+1$)-dimensional regular polyhedron with a distance between vertices equal to $r_0$. The special properties of this class of models result from the fact that all domain walls have the same tension, so that only Y-type junctions are allowed. Also, the probability of fusion of adjacent domains as a result of domain wall collapse decreases proportionally to $1/N_\phi$ for large $N_\phi$, thus vanishing in the $N_\phi \to \infty$ limit.

Although other classes of models allowing for X-type junctions \cite{Bazeia:2005wt} have also been investigated in Refs. \cite{Avelino:2008ve,Avelino:2006xy} it has been argued that the ideal model was the best model from the point of view of its potential to lead to frustration, a necessary condition for domain walls to provide a significant contribution to dark energy. On the other hand, in a non-relativistic friction-dominated regime and in the absence of expansion, the ideal model can be used to describe the dynamics of ideal grain growth. 

Here we compare the results of a set of high resolution interface network simulations of the ideal class of models with $N_\phi=10$ ($11$ minima represented by $11$ different colors in Fig.~1), considering the relativistic and non-relativistic regimes studied in the previous section (we have taken $\alpha=2/3$ and $\alpha=0$, respectively). Fig.~1 shows two $256^2$ snapshots taken from relativistic (left) and nonrelativistic (right) $1024^2$ two-dimensional interface simulations ($\partial \phi/\partial z=0$) with periodic boundaries and identical initial conditions: At each grid point $\phi$ was randomly assigned to one of the minima of the potential and $\dot \phi$ was initially set to zero (see Ref. \cite{Avelino:2008ve} for a more detailed account of simulation details). The characteristic length of the network is roughly the same in both snapshots (note that the corresponding physical times can be very different). Fig.~1 shows that although relativistic and nonrelativistic interface networks have very different dynamics, a single simulation snapshot does not clearly distinguish between the two regimes (we verified that this result is valid irrespective of the specific choice of $N_\phi$).

This is shown more quantitatively in Fig.~2, which summarizes the results of ten relativistic and nonrelativistic snapshots with $1024$ domains each, taken from $1024^2$ numerical simulations. Again, the same initial conditions were considered for each pair of relativistic-nonrelativistic simulations. The blue stripe represents the interval $f_n \pm s_n$ where $f_n$ is the fraction of the total number of domains with $n$ edges for either relativistic (upper panel) or nonrelativistic (lower panel) regimes and $s_n^2$ is the sample variance calculated using ten $1024^2$ domain wall network simulations. The red dots represent the value of $f_n$ in the nonrelativistic (upper panel) or relativistic cases (lower panel) for comparison. We verified that the average number of edges of an individual domain, numerically calculated from the simulations, is equal to $6$ in both cases, a characteristic of interface networks with Y-type junctions.  Also, we observed no significant changes to $f_n$ on subsequent snapshots.

\begin{figure}[t!]
  \includegraphics[width=8.5cm]{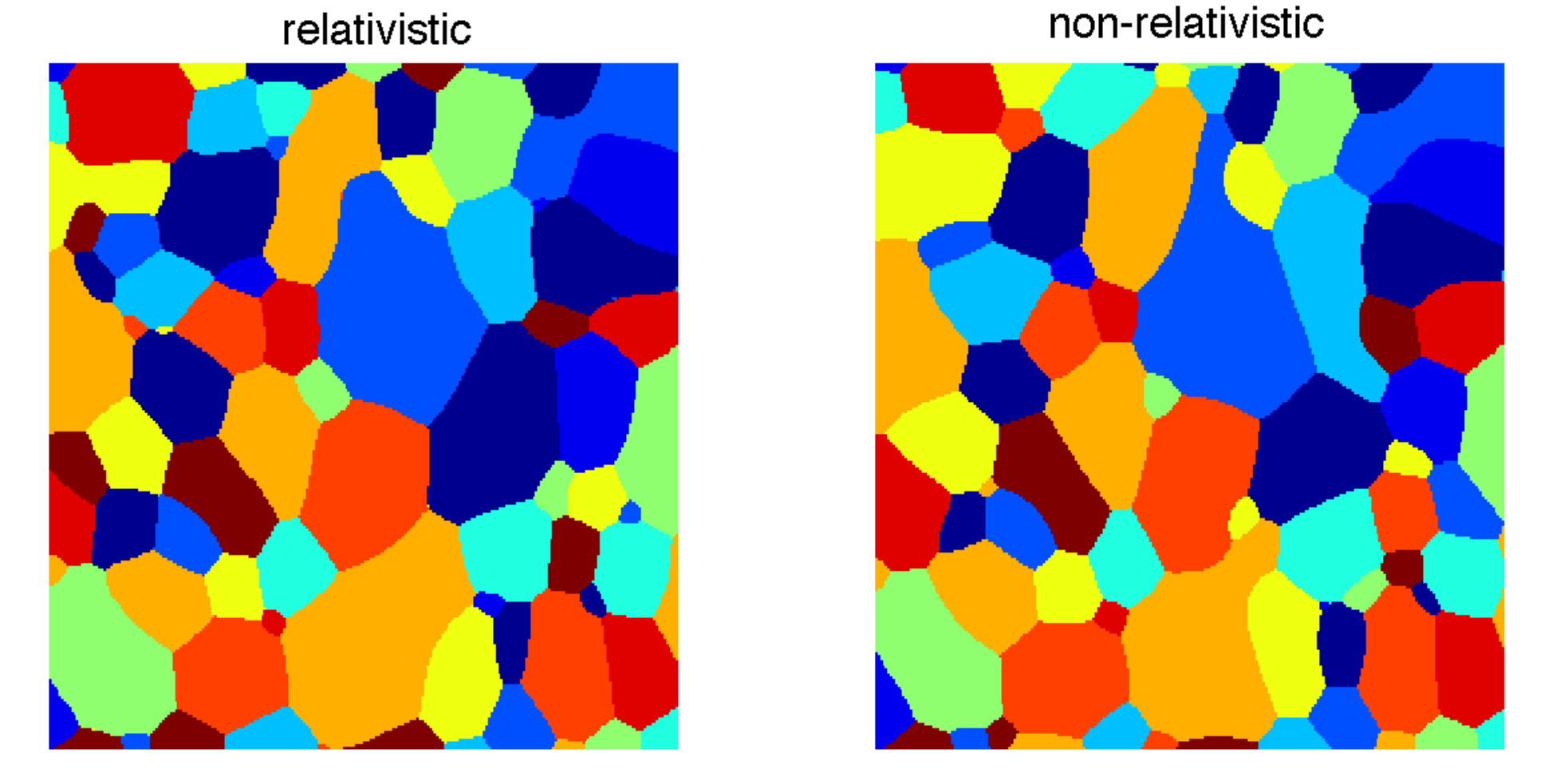}
  \caption{\small{(Color online) Two $256^2$ simulation snapshots, with similar characteristic lengths, taken from $1024^2$ relativistic (left) and nonrelativistic (right) two-dimensional interface simulations of the ideal class of models with $N_\phi=10$, with identical initial conditions. Different colors represent different minima.}}
\end{figure}

The distribution function $f_n$ has been extensively used to characterize nonrelativistic two-dimensional patterns (see Ref.  \cite{1992JPCM....4.1867G} for a review). Analytical predictions for $f_n$ have been made in the context of soap froth and other similar systems \cite{0295-5075-15-5-002,PhysRevLett.69.2674,PhysRevE.47.4037}. In particular, the scale-invariant distribution function $f_n$ for two-dimensional soap froths \cite{PhysRevLett.62.1318} was found to be in very good agreement with the Monte Carlo simulations of the ideal Potts model  \cite{GGA}, corresponding to an infinite number of different domains. The most significant difference between the nonrelativistic simulation results of Fig.~2 (with $N_\phi=10$) and soap froth experimental results is that no two-sided domains are actually observed in real soap froths (corresponding to $N_\phi=\infty$). The fact that $f_n$ is maximal for $n=5$ is rather standard in relaxed two-dimensional systems with Y-type junctions, such as real soap froths.

It was demonstrated in Ref. \cite{Avelino:2008ve} that in order to provide a significant contribution to the dark energy without generating exceedingly large CMB temperature fluctuations, a domain wall network must have a characteristic length $ L \lsim 10  \, {\rm kpc}$ and a characteristic velocity ${\bar v} \lsim 10^{-5}$ at the present time. That would be possible only if the relativistic distribution function in Fig.~2 was sharply peaked around $n=6$. As shown in Fig.~2 this is clearly not the case. Hence, the presence of two-, three-, four-, and five-edge unstable domains is enough to prevent the frustration of the interface network. 

\begin{figure}[t!]
\includegraphics[width=8.5cm]{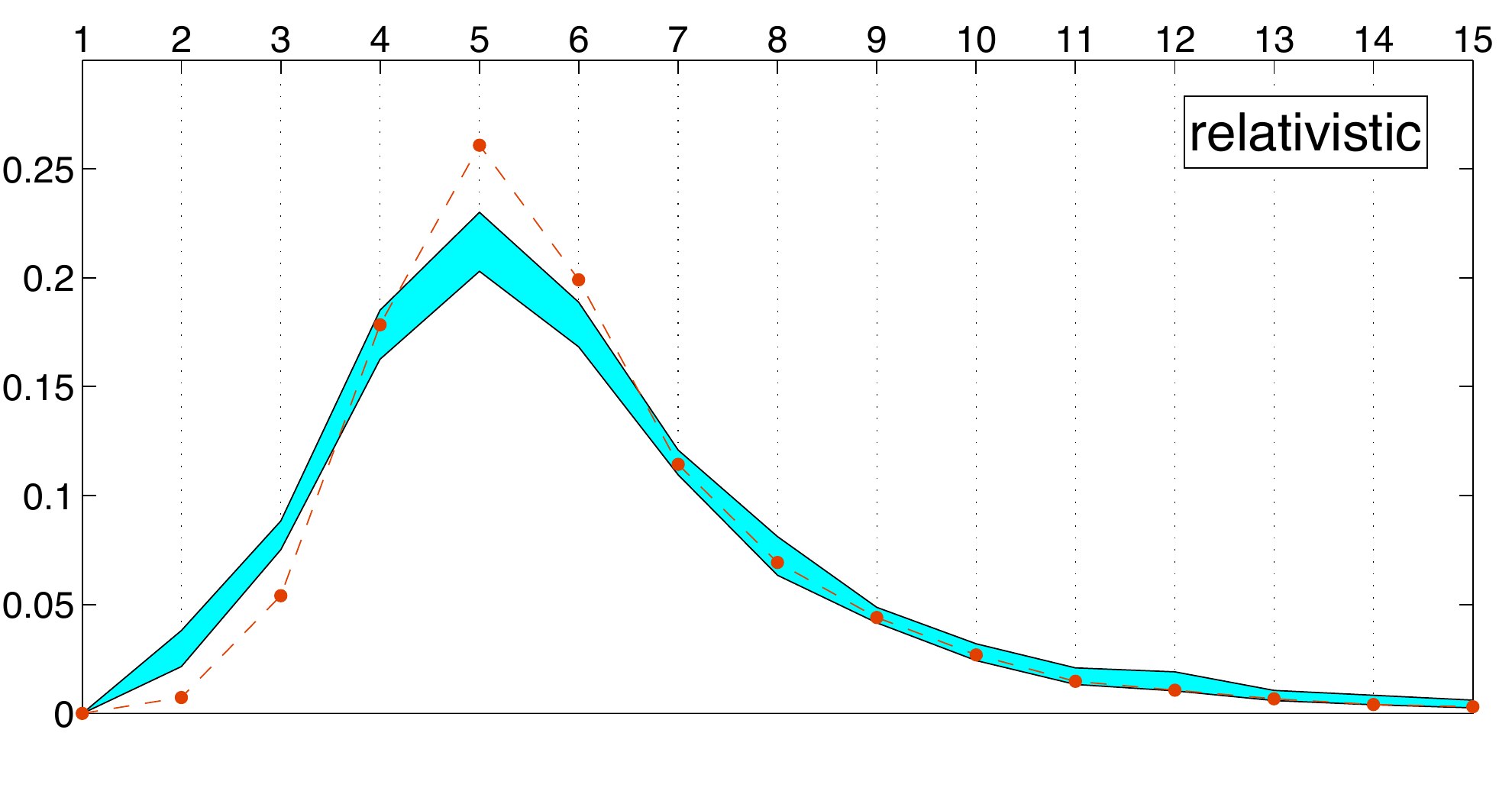}
\includegraphics[width=8.5cm]{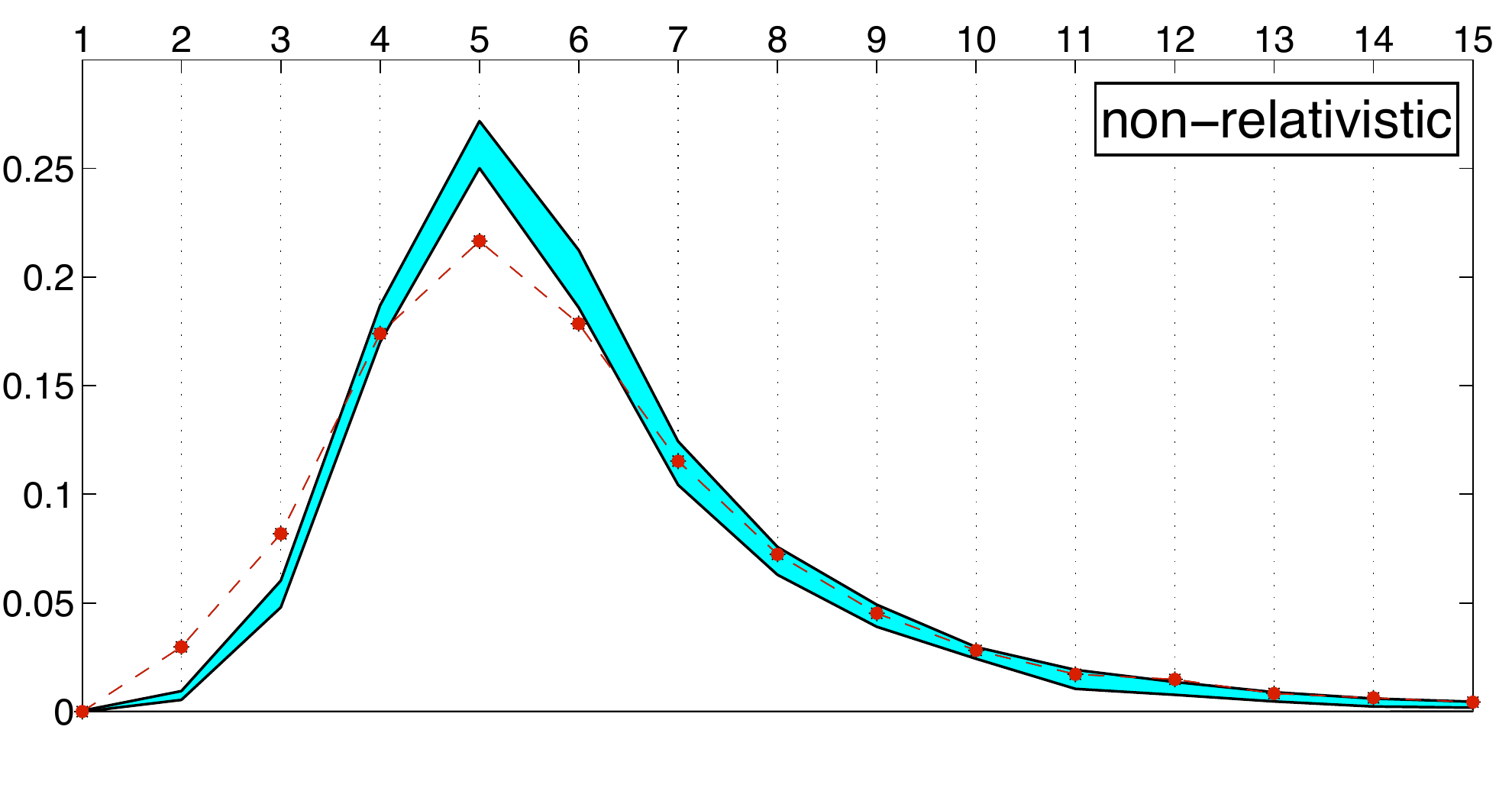}
\vspace{-0.5cm}
\caption{\small{(Color online) The blue (light gray) stripe represents the interval $f_n \pm s_n$, where $f_n$ is the fraction of the number of domains with $n$ edges for either relativistic (upper panel) or nonrelativistic (lower panel) regimes and $s_n^2$ is the sample variance calculated using ten snapshots with $1024$ domains each, taken from $1024^2$ numerical simulations of the ideal class of models with $N_\phi=10$. The red dots represent the value of $f_n$ in the nonrelativistic (upper panel) or relativistic cases (lower panel) for comparison.}}
\end{figure}

Fig.~2 shows that for $n>6$ the number density of domains with a given number of edges is similar for relativistic and nonrelativistic networks, in agreement with  Fig.~1, although there are some differences for $n \le 6$. In the relativistic case the velocity can never be much larger than ${\bar v}$, since it is limited by the speed of light. On the other hand, in the nonrelativistic regime, $v \propto \kappa \ell_{\rm f} \gg {\bar v}$ for highly curved interfaces, leading to the suppression of very small domains in the nonrelativistic regime. This can be seen in Fig.~1, and it is the main source of the differences between the relativistic and nonrelativistic results shown in Fig.~2. Still, the similarity between relativistic and nonrelativistic snapshots of simulations with identical initial conditions is striking, thus demonstrating that it is the initial conditions rather than the specific dynamics of the cellular pattern which essentially determine the large scale network properties at late times. Although in this paper only the results of interface simulations in two spatial dimensions are presented, we have verified that similar results also hold in the case of two-dimensional snapshots of three-dimensional interface network simulations.

These results suggest that a modified statistical version of von Neumann' s law might be valid for relativistic interface networks. This can be confirmed by determining the evolution of the comoving area of a single domain of $n$ sides with physical time, ${\dot a}_n=-R^{-1}\oint v dl$, where $dl$ is the comoving infinitesimal interface arc length (note that the physical length and area may be obtained by multiplying the comoving ones by $R$ and $R^2$, respectively). Ignoring the ${\dot v}$ term in  Eq. (\ref{dyn1}) one has
\be
{\dot a}_n= - \frac{\ell_{\rm d}}{R^2}\oint \kappa_c dl =- \frac{\ell_{\rm d}}{R^2} \left(2\pi - \sum_{i=1}^n \theta_i\right)\,,
\ee
where $k_c=R k$ and $\theta_i$ represent each of the $n$ discontinuous angle changes at the vertices. If the interface network has only Y-type junctions which meet at an angle of $2 \pi/3$, then
\be
{\dot a}_n=-\frac{\ell_{\rm d}}{R^2} \left[2\pi - n\left(\pi-\frac{2\pi}{3}\right)\right]=\frac{\pi}{3} \frac{\ell_{\rm d}}{R^2}  (n-6)\,.
\label{dandt}
\ee
For scaling relativistic interface networks it is ${\dot {\bar v}}$ and not ${\dot v}$ which is equal to zero and consequently von Neumann's law is valid only on average. The evolution of the total number of domains, $N_D$, with time can also be obtained from Eq. (\ref{dandt}). If ${\dot f}_n=0$ then
\be
\label{dntdt}
\frac{{\dot N}_D}{N_D}= \sum_{n < {\bar n}} \left(\frac{{\dot a}_n}{{\bar a}_n} f_n\right)\,.
\ee
Here $A=\sum_{n=1}^\infty A_n$ is the comoving area of the entire system, $A_n$ is the total comoving area occupied by domains with $n$ edges, ${\bar a}=A/N_D$ is the average comoving domain area, and ${\bar a}_n=g_n {\bar a}$ is the average comoving area of a domain with $n$ edges. If both $f_n$ and $g_n$ are time independent, then ${\dot N}_D \propto  -\ell_{\rm d} N_D^2/R^2$.  Eqs. (\ref{dandt}) and (\ref{dntdt}) imply that  $L \propto R N_D^{-1/2}$ is proportional to $t$  in the frictionless relativistic case ($\ell_{\rm d}^{-1} \propto H=\alpha/t$, with $0 <\alpha<1$) and proportional to $t^{1/2}$ in the nonrelativistic friction dominated regime ($\ell_{\rm d}=\ell_{\rm f}$ and $R=1$), in agreement with Eqs. (\ref{scaling}) and (\ref{scalingnr}). Therefore, the main difference between relativistic and nonrelativistic thin featureless interface networks with the same ${\mathcal L}$ and initial conditions is associated with the scalings of the macroscopic parameters ${\bar v}$ and $L$ with physical time, $t$, rather than with the properties of the cellular patterns taken from single simulation snapshots with a fixed $L$. Given that frustration, a necessary condition for a domain wall network to be able to accelerate the expansion of the universe, cannot happen if there are unstable domains with less than six edges, laboratory tests, such as those probing the dynamics of real foam coarsening \cite{1998PhRvL..80.5228M} (see also  Ref. \cite{foams} and references therein) can be used to rule out domain wall networks as a relevant dark energy source.

\section{Conclusion}

In this paper we have provided a rigorous derivation of the VOS domain wall model, showing that it may be used to successfully describe the dynamics of both relativistic and nonrelativistic interfaces. Although the evolution of the characteristic length and velocity of interface networks  with physical time is very different in these regimes, we have shown that a single simulation snapshot does not clearly distinguish relativistic and nonrelativistic networks with the same characteristic length, the reason being that a modified version of von Neumann's law also applies (in a statistical sense) to the relativistic regime. We have argued that this allows the use of laboratory tests with nonrelativistic interfaces to rule out cosmological domain wall networks as a significant dark energy source.

\begin{acknowledgments}

The authors would like to thank CAPES, CNPq (Brazil), and FCT (Portugal) for partial support.

\end{acknowledgments}

\bibliography{interfaces}

\begin{thebibliography}{27}
\expandafter\ifx\csname natexlab\endcsname\relax\def\natexlab#1{#1}\fi
\expandafter\ifx\csname bibnamefont\endcsname\relax
  \def\bibnamefont#1{#1}\fi
\expandafter\ifx\csname bibfnamefont\endcsname\relax
  \def\bibfnamefont#1{#1}\fi
\expandafter\ifx\csname citenamefont\endcsname\relax
  \def\citenamefont#1{#1}\fi
\expandafter\ifx\csname url\endcsname\relax
  \def\url#1{\texttt{#1}}\fi
\expandafter\ifx\csname urlprefix\endcsname\relax\def\urlprefix{URL }\fi
\providecommand{\bibinfo}[2]{#2}
\providecommand{\eprint}[2][]{\url{#2}}

\bibitem[{\citenamefont{{Glazier} and {Weaire}}(1992)}]{1992JPCM....4.1867G}
\bibinfo{author}{\bibfnamefont{J.~A.} \bibnamefont{{Glazier}}}
  \bibnamefont{and} \bibinfo{author}{\bibfnamefont{D.}~\bibnamefont{{Weaire}}},
  \bibinfo{journal}{Journal of Physics Condensed Matter}
  \textbf{\bibinfo{volume}{4}}, \bibinfo{pages}{1867} (\bibinfo{year}{1992}).

\bibitem[{\citenamefont{Weaire and Hutzler}(2000)}]{foams}
\bibinfo{author}{\bibfnamefont{D.}~\bibnamefont{Weaire}} \bibnamefont{and}
  \bibinfo{author}{\bibfnamefont{R.}~\bibnamefont{Hutzler}},
  \emph{\bibinfo{title}{The physics of foams}} (\bibinfo{publisher}{Oxford
  University Press}, \bibinfo{address}{Oxford}, \bibinfo{year}{2000}).

\bibitem[{\citenamefont{{Kim} et~al.}(2006)\citenamefont{{Kim}, {Kim}, {Kim},
  and {Park}}}]{2006PhRvE..74f1605K}
\bibinfo{author}{\bibfnamefont{S.~G.} \bibnamefont{{Kim}}},
  \bibinfo{author}{\bibfnamefont{D.~I.} \bibnamefont{{Kim}}},
  \bibinfo{author}{\bibfnamefont{W.~T.} \bibnamefont{{Kim}}}, \bibnamefont{and}
  \bibinfo{author}{\bibfnamefont{Y.~B.} \bibnamefont{{Park}}},
  \bibinfo{journal}{\pre} \textbf{\bibinfo{volume}{74}},
  \bibinfo{pages}{061605} (\bibinfo{year}{2006}).

\bibitem[{\citenamefont{{Bray}}(2002)}]{2002AdPhy..51..481B}
\bibinfo{author}{\bibfnamefont{A.~J.} \bibnamefont{{Bray}}},
  \bibinfo{journal}{Advances in Physics} \textbf{\bibinfo{volume}{51}},
  \bibinfo{pages}{481} (\bibinfo{year}{2002}).

\bibitem[{\citenamefont{{Emmerich}}(2008)}]{2008AdPhy..57....1E}
\bibinfo{author}{\bibfnamefont{H.}~\bibnamefont{{Emmerich}}},
  \bibinfo{journal}{Advances in Physics} \textbf{\bibinfo{volume}{57}},
  \bibinfo{pages}{1} (\bibinfo{year}{2008}).

\bibitem[{\citenamefont{{Chen} and {Yang}}(1994)}]{1994PhRvB..5015752C}
\bibinfo{author}{\bibfnamefont{L.}~\bibnamefont{{Chen}}} \bibnamefont{and}
  \bibinfo{author}{\bibfnamefont{W.}~\bibnamefont{{Yang}}},
  \bibinfo{journal}{\prb} \textbf{\bibinfo{volume}{50}}, \bibinfo{pages}{15752}
  (\bibinfo{year}{1994}).

\bibitem[{\citenamefont{{Monnereau} and
  {Vignes-Adler}}(1998)}]{1998PhRvL..80.5228M}
\bibinfo{author}{\bibfnamefont{C.}~\bibnamefont{{Monnereau}}} \bibnamefont{and}
  \bibinfo{author}{\bibfnamefont{M.}~\bibnamefont{{Vignes-Adler}}},
  \bibinfo{journal}{Physical Review Letters} \textbf{\bibinfo{volume}{80}},
  \bibinfo{pages}{5228} (\bibinfo{year}{1998}).

\bibitem[{\citenamefont{von Neumann}(1952)}]{vonNeumann:1952:DRC}
\bibinfo{author}{\bibfnamefont{J.}~\bibnamefont{von Neumann}}, in
  \emph{\bibinfo{booktitle}{Metal Interfaces}} (\bibinfo{year}{1952}), pp.
  \bibinfo{pages}{108--110}.

\bibitem[{\citenamefont{Stavans et~al.}(1991)\citenamefont{Stavans, Domany, and
  Mukamel}}]{0295-5075-15-5-002}
\bibinfo{author}{\bibfnamefont{J.}~\bibnamefont{Stavans}},
  \bibinfo{author}{\bibfnamefont{E.}~\bibnamefont{Domany}}, \bibnamefont{and}
  \bibinfo{author}{\bibfnamefont{D.}~\bibnamefont{Mukamel}},
  \bibinfo{journal}{EPL (Europhysics Letters)} \textbf{\bibinfo{volume}{15}},
  \bibinfo{pages}{479} (\bibinfo{year}{1991}).

\bibitem[{\citenamefont{{Avron} and {Levine}}(1992)}]{1992PhRvL..69..208A}
\bibinfo{author}{\bibfnamefont{J.~E.} \bibnamefont{{Avron}}} \bibnamefont{and}
  \bibinfo{author}{\bibfnamefont{D.}~\bibnamefont{{Levine}}},
  \bibinfo{journal}{Physical Review Letters} \textbf{\bibinfo{volume}{69}},
  \bibinfo{pages}{208} (\bibinfo{year}{1992}).

\bibitem[{\citenamefont{Sire}(1994)}]{PhysRevLett.72.420}
\bibinfo{author}{\bibfnamefont{C.}~\bibnamefont{Sire}}, \bibinfo{journal}{Phys.
  Rev. Lett.} \textbf{\bibinfo{volume}{72}}, \bibinfo{pages}{420}
  (\bibinfo{year}{1994}).

\bibitem[{\citenamefont{{Hilgenfeldt} et~al.}(2001)\citenamefont{{Hilgenfeldt},
  {Kraynik}, {Koehler}, and {Stone}}}]{2001PhRvL..86.2685H}
\bibinfo{author}{\bibfnamefont{S.}~\bibnamefont{{Hilgenfeldt}}},
  \bibinfo{author}{\bibfnamefont{A.~M.} \bibnamefont{{Kraynik}}},
  \bibinfo{author}{\bibfnamefont{S.~A.} \bibnamefont{{Koehler}}},
  \bibnamefont{and} \bibinfo{author}{\bibfnamefont{H.~A.}
  \bibnamefont{{Stone}}}, \bibinfo{journal}{Physical Review Letters}
  \textbf{\bibinfo{volume}{86}}, \bibinfo{pages}{2685} (\bibinfo{year}{2001}).

\bibitem[{\citenamefont{{MacPherson,} and {Srolovitz,}}(2007)}]{nature1}
\bibinfo{author}{\bibfnamefont{R.~D.} \bibnamefont{{MacPherson,}}}
  \bibnamefont{and} \bibinfo{author}{\bibfnamefont{D.~J.}
  \bibnamefont{{Srolovitz,}}}, \bibinfo{journal}{Nature}
  \textbf{\bibinfo{volume}{446}}, \bibinfo{pages}{1053} (\bibinfo{year}{2007}).

\bibitem[{\citenamefont{{Weaire} and {Rivier}}(2009)}]{2009ConPh..50..199W}
\bibinfo{author}{\bibfnamefont{D.}~\bibnamefont{{Weaire}}} \bibnamefont{and}
  \bibinfo{author}{\bibfnamefont{N.}~\bibnamefont{{Rivier}}},
  \bibinfo{journal}{Contemporary Physics} \textbf{\bibinfo{volume}{50}},
  \bibinfo{pages}{199} (\bibinfo{year}{2009}).

\bibitem[{\citenamefont{Hilhorst}(2006)}]{0305-4470-39-23-004}
\bibinfo{author}{\bibfnamefont{H.~J.} \bibnamefont{Hilhorst}},
  \bibinfo{journal}{Journal of Physics A: Mathematical and General}
  \textbf{\bibinfo{volume}{39}}, \bibinfo{pages}{7227} (\bibinfo{year}{2006}).

\bibitem[{\citenamefont{Flyvbjerg}(1993)}]{PhysRevE.47.4037}
\bibinfo{author}{\bibfnamefont{H.}~\bibnamefont{Flyvbjerg}},
  \bibinfo{journal}{Phys. Rev. E} \textbf{\bibinfo{volume}{47}},
  \bibinfo{pages}{4037} (\bibinfo{year}{1993}).

\bibitem[{\citenamefont{Bucher and Spergel}(1999)}]{Bucher:1998mh}
\bibinfo{author}{\bibfnamefont{M.}~\bibnamefont{Bucher}} \bibnamefont{and}
  \bibinfo{author}{\bibfnamefont{D.~N.} \bibnamefont{Spergel}},
  \bibinfo{journal}{Phys. Rev.} \textbf{\bibinfo{volume}{D60}},
  \bibinfo{pages}{043505} (\bibinfo{year}{1999}).

\bibitem[{\citenamefont{Avelino
  et~al.}(2006{\natexlab{a}})\citenamefont{Avelino, Martins, Menezes, Menezes,
  and Oliveira}}]{PinaAvelino:2006ia}
\bibinfo{author}{\bibfnamefont{P.~P.} \bibnamefont{Avelino}},
  \bibinfo{author}{\bibfnamefont{C.~J. A.~P.} \bibnamefont{Martins}},
  \bibinfo{author}{\bibfnamefont{J.}~\bibnamefont{Menezes}},
  \bibinfo{author}{\bibfnamefont{R.}~\bibnamefont{Menezes}}, \bibnamefont{and}
  \bibinfo{author}{\bibfnamefont{J.~C. R.~E.} \bibnamefont{Oliveira}},
  \bibinfo{journal}{Phys. Rev.} \textbf{\bibinfo{volume}{D73}},
  \bibinfo{pages}{123519} (\bibinfo{year}{2006}{\natexlab{a}}).

\bibitem[{\citenamefont{Avelino et~al.}(2008)\citenamefont{Avelino, Martins,
  Menezes, Menezes, and Oliveira}}]{Avelino:2008ve}
\bibinfo{author}{\bibfnamefont{P.~P.} \bibnamefont{Avelino}},
  \bibinfo{author}{\bibfnamefont{C.~J. A.~P.} \bibnamefont{Martins}},
  \bibinfo{author}{\bibfnamefont{J.}~\bibnamefont{Menezes}},
  \bibinfo{author}{\bibfnamefont{R.}~\bibnamefont{Menezes}}, \bibnamefont{and}
  \bibinfo{author}{\bibfnamefont{J.~C. R.~E.} \bibnamefont{Oliveira}},
  \bibinfo{journal}{Phys. Rev.} \textbf{\bibinfo{volume}{D78}},
  \bibinfo{pages}{103508} (\bibinfo{year}{2008}).

\bibitem[{\citenamefont{Sousa and Avelino}(2010)}]{Sousa:2010zz}
\bibinfo{author}{\bibfnamefont{L.}~\bibnamefont{Sousa}} \bibnamefont{and}
  \bibinfo{author}{\bibfnamefont{P.~P.} \bibnamefont{Avelino}},
  \bibinfo{journal}{Phys. Rev.} \textbf{\bibinfo{volume}{D81}},
  \bibinfo{pages}{087305} (\bibinfo{year}{2010}).

\bibitem[{\citenamefont{Martins and Shellard}(2002)}]{Martins:2000cs}
\bibinfo{author}{\bibfnamefont{C.~J. A.~P.} \bibnamefont{Martins}}
  \bibnamefont{and} \bibinfo{author}{\bibfnamefont{E.~P.~S.}
  \bibnamefont{Shellard}}, \bibinfo{journal}{Phys. Rev.}
  \textbf{\bibinfo{volume}{D65}}, \bibinfo{pages}{043514}
  (\bibinfo{year}{2002}).

\bibitem[{\citenamefont{Avelino et~al.}(2005)\citenamefont{Avelino, Martins,
  and Oliveira}}]{Avelino:2005kn}
\bibinfo{author}{\bibfnamefont{P.~P.} \bibnamefont{Avelino}},
  \bibinfo{author}{\bibfnamefont{C.~J. A.~P.} \bibnamefont{Martins}},
  \bibnamefont{and} \bibinfo{author}{\bibfnamefont{J.~C. R.~E.}
  \bibnamefont{Oliveira}}, \bibinfo{journal}{Phys. Rev.}
  \textbf{\bibinfo{volume}{D72}}, \bibinfo{pages}{083506}
  (\bibinfo{year}{2005}).

\bibitem[{\citenamefont{Bazeia et~al.}(2006)\citenamefont{Bazeia, Brito, and
  Losano}}]{Bazeia:2005wt}
\bibinfo{author}{\bibfnamefont{D.}~\bibnamefont{Bazeia}},
  \bibinfo{author}{\bibfnamefont{F.~A.} \bibnamefont{Brito}}, \bibnamefont{and}
  \bibinfo{author}{\bibfnamefont{L.}~\bibnamefont{Losano}},
  \bibinfo{journal}{Europhys. Lett.} \textbf{\bibinfo{volume}{D76}},
  \bibinfo{pages}{374} (\bibinfo{year}{2006}).

\bibitem[{\citenamefont{Avelino
  et~al.}(2006{\natexlab{b}})\citenamefont{Avelino, Martins, Menezes, Menezes,
  and Oliveira}}]{Avelino:2006xy}
\bibinfo{author}{\bibfnamefont{P.~P.} \bibnamefont{Avelino}},
  \bibinfo{author}{\bibfnamefont{C.~J. A.~P.} \bibnamefont{Martins}},
  \bibinfo{author}{\bibfnamefont{J.}~\bibnamefont{Menezes}},
  \bibinfo{author}{\bibfnamefont{R.}~\bibnamefont{Menezes}}, \bibnamefont{and}
  \bibinfo{author}{\bibfnamefont{J.~C. R.~E.} \bibnamefont{Oliveira}},
  \bibinfo{journal}{Phys. Rev.} \textbf{\bibinfo{volume}{D73}},
  \bibinfo{pages}{123520} (\bibinfo{year}{2006}{\natexlab{b}}),
  \eprint{hep-ph/0604250}.

\bibitem[{\citenamefont{Godr\`eche et~al.}(1992)\citenamefont{Godr\`eche,
  Kostov, and Yekutieli}}]{PhysRevLett.69.2674}
\bibinfo{author}{\bibfnamefont{C.}~\bibnamefont{Godr\`eche}},
  \bibinfo{author}{\bibfnamefont{I.}~\bibnamefont{Kostov}}, \bibnamefont{and}
  \bibinfo{author}{\bibfnamefont{I.}~\bibnamefont{Yekutieli}},
  \bibinfo{journal}{Phys. Rev. Lett.} \textbf{\bibinfo{volume}{69}},
  \bibinfo{pages}{2674} (\bibinfo{year}{1992}).

\bibitem[{\citenamefont{Stavans and Glazier}(1989)}]{PhysRevLett.62.1318}
\bibinfo{author}{\bibfnamefont{J.}~\bibnamefont{Stavans}} \bibnamefont{and}
  \bibinfo{author}{\bibfnamefont{J.~A.} \bibnamefont{Glazier}},
  \bibinfo{journal}{Phys. Rev. Lett.} \textbf{\bibinfo{volume}{62}},
  \bibinfo{pages}{1318} (\bibinfo{year}{1989}).

\bibitem[{\citenamefont{Glazier et~al.}(1990)\citenamefont{Glazier, Grest, and
  Anderson}}]{GGA}
\bibinfo{author}{\bibfnamefont{A.}~\bibnamefont{Glazier}},
  \bibinfo{author}{\bibfnamefont{G.}~\bibnamefont{Grest}}, \bibnamefont{and}
  \bibinfo{author}{\bibfnamefont{M.}~\bibnamefont{Anderson}}, in
  \emph{\bibinfo{booktitle}{Simulation and Theory of Evolving
  Microstructures}}, edited by
  \bibinfo{editor}{\bibfnamefont{M.}~\bibnamefont{Anderson}} \bibnamefont{and}
  \bibinfo{editor}{\bibfnamefont{A.}~\bibnamefont{Rollett}}
  (\bibinfo{publisher}{The Minerals, Metals and Materials Society},
  \bibinfo{year}{1990}).

\end{thebibliography}

\end{document}